\documentclass[letterpaper,preprint,byrevtex,showpacs,groupedaddress,pre]{revtex4-1}

\DeclareSymbolFont{lettersA}{U}{pxmia}{m}{it}
\DeclareMathAlphabet{\mathsfsl}{OT1}{cmss}{m}{sl}

\SetSymbolFont{lettersA}{bold}{U}{pxmia}{bx}{it}
\DeclareFontSubstitution{U}{pxmia}{m}{it}
\DeclareSymbolFontAlphabet{\mathfrak}{lettersA}
\DeclareMathSymbol{\piup}{\mathord}{lettersA}{"19}
\DeclareMathSymbol{\iTheta}{\mathalpha}{letters}{2}

\usepackage{amsmath}
\usepackage{amssymb}
\usepackage{graphicx}
\usepackage{bbm}
\usepackage{mathrsfs}
\usepackage{xcolor}
\usepackage[colorlinks=true,pdfstartview=FitV,linkcolor=blue,citecolor=blue,urlcolor=blue]{hyperref}

\newcommand{\ntensor}[1]{\mathord{\buildrel{\lower3pt\hbox{$\scriptscriptstyle\leftrightarrow$}}\over{#1}}}

\makeatletter

\newcommand{\Rmnum}[1]{\expandafter\@slowromancap\romannumeral #1@}
\makeatother

\begin{document}


\title[Light propagation in extremely anisotropic metamaterials]
{Diffraction-free optical beam propagation with near-zero phase variation in extremely anisotropic metamaterials}

\author{Lei~Sun}
\affiliation{Department of Mechanical and Aerospace Engineering, \\
    Missouri University of Science and Technology, \\
    Rolla, Missouri 65409, USA}

\author{Xiaodong~Yang}
\email[Electronic address: ]{yangxia@mst.edu}
\affiliation{Department of Mechanical and Aerospace Engineering, \\
    Missouri University of Science and Technology, \\
    Rolla, Missouri 65409, USA}

\author{Wei~Wang}
\affiliation{Department of Mechanical and Aerospace Engineering, \\
    Missouri University of Science and Technology, \\
    Rolla, Missouri 65409, USA}

\author{Jie~Gao}
\email[Electronic address: ]{gaojie@mst.edu}
\affiliation{Department of Mechanical and Aerospace Engineering, \\
    Missouri University of Science and Technology, \\
    Rolla, Missouri 65409, USA}

%
%

\begin{abstract}
Extremely anisotropic metal-dielectric multilayer metamaterials are designed
to have the effective permittivity tensor of a transverse component
(parallel to the interfaces of the multilayer) with zero real part and
a longitudinal component (normal to the interfaces of the multilayer)
with ultra-large imaginary part at the same wavelength, including the optical
nonlocality analysis based on the transfer-matrix method.
The diffraction-free deep-subwavelength optical beam propagation with near-zero
phase variation in the designed multilayer stack due to the near-flat iso-frequency
contour is demonstrated and analyzed, including the effects of the multilayer
period and the material loss.
\end{abstract}


\maketitle

\section{Introduction}

Light beam propagating beyond diffraction limit with deep-subwavelength confinement
and near-zero phase variation is highly desirable in many optical integration applications.
In order to realize such intriguing optical beam propagation phenomena, optical materials
are required to possess an extremely anisotropic effective refractive index tensor
with an ultra-high component perpendicular to the propagation direction to support the light
beam of large wave vector and an ultra-low component in the propagation direction to
reduce the phase variation of the light beam simultaneously.
Optical metamaterials with artificial subwavelength meta-atoms and tunable effective
permittivity and effective permeability have been designed to exhibit ultra-high
effective refractive index so as to achieve large wave vectors \cite{Shin2009PRL,Choi2011Nat},
as well as to have zero effective refractive index for obtaining the optical beam propagation
with zero phase variation \cite{Huang2011NatMate,Vesseur2013PRL}.
Among all these designs, the metal-dielectric multilayer metamaterials are of great interest
in demonstrating many exciting applications
such as negative refraction \cite{Valentine2008Nat},
epsilon-near-zero (ENZ) materials \cite{Maas2013NatPhoto,Gao2013APL,Sun2013OE,Sun2014OE},
and epsilon-near-pole (ENP) materials \cite{Molesky2013OE}.
Particularly, possessing an indefinite permittivity tensor, the multilayer metamaterials
are of unique applications in
subwavelength imaging \cite{Liu2007Sci,Zhang2008NatMate},
enhanced photonic density of states \cite{Krishnamoorthy2012Sci},
broadband light absorbers \cite{Cui2012NanoLett,Guclu2012PRB,Narimanov2013OE},
ultra-high refractive indices for subwavelength optical waveguides \cite{He2013JOSAB},
indefinite cavities \cite{Yao2011PNAS,Yang2012NatPhoto},
broadband spontaneous emission engineering \cite{Noginov2010OL,Kidwai2011OL,Jacob2012APL},
super-Plank thermal emission \cite{Biehs2012PRL,Guo2012APL,Guo2013OE},
ultra-large-wavevector plasmon polaritons \cite{Othman2013JN,Zhukovsky2013OE,Zhukovsky2014PRB},
and other exceptional properties \cite{Cortes2012JO,Poddubny2013NP,Orlova2014PN}.

In our previous work \cite{He2013JO}, the extremely loss-anisotropic metal-dielectric multilayer
metamaterials are designed in order to realize the diffraction-free optical beam
propagation mainly based on the effective medium theory (EMT) and the operation wavelength is adjusted from the
dispersion relation of the multilayer stack.
However, the effective permittivity tensor including
the optical nonlocality is not well considered in the analysis.
Therefore, in the present work, a new kind of metal-dielectric multilayer metamaterials are
designed to have the extremely anisotropic effective permittivity tensor for
the demonstration of both the diffraction-free optical beam propagation and
the near-zero phase variation along the propagation direction, based on a
theoretical analysis of the effective permittivity tensor including the optical nonlocality.
The designed multilayer metamaterials will have the effective permittivity tensor
with a near-zero transverse component parallel to the multilayer and an ultra-large
longitudinal component in the imaginary part normal to the multilayer at the same
wavelength, which represents the ENZ-ENP wavelength.
The near-zero transverse component of the effective permittivity tensor leads to
the near-zero phase variation during the propagation of the electromagnetic wave.
On the other hand, the longitudinal component with an ultra-large imaginary part
of the effective permittivity tensor will reduce the diffraction of the electromagnetic
wave \cite{Feng2012PRL,Sun2012APL}.
The filling ratio of metal in the metal-dielectric multilayer stack with respect to
the ENZ-ENP wavelength is fully analyzed according to the transfer-matrix method
including the optical nonlocality, while the performance of the deep-subwavelength
optical beam propagation with near-zero phase variation is also explored with respect
to the variation of the material loss.
In the following sections, the diffraction-free optical beam propagation with
near-zero phase variation is discussed in theoretical analysis based on
EMT and optical nonlocality, together with numerical
simulation in Sec.~2, followed by the conclusion in Sec.~3.

\section{Theory and discussion}

\subsection{Effective medium theory and optical nonlocality}
Illustrated in Fig.~1(a), the metal-dielectric multilayer stack is composite of alternating
layers of silver ($\mathrm{Ag}$) and titanium pentoxide ($\mathrm{Ti}_{3}\mathrm{O}_{5}$),
with the thickness $a_{m}$ and $a_{d}$, respectively.
The permittivity of $\mathrm{Ag}$ follows the Drude model
$\varepsilon_{m}=\varepsilon_{\infty}-\omega_{p}^{2}/\left( \omega^{2}+i\omega\gamma \right)$,
with the permittivity constant $\varepsilon_{\infty}=5.7$,
the plasma frequency $\omega_{p}=1.37\times10^{16}\,\mathrm{rad}/\mathrm{s}$,
and the damping factor $\gamma=\Delta\cdot(8.5\times10^{13})\,\mathrm{rad}/\mathrm{s}$.
According to the previous experimental demonstration \cite{Chen2010OE},
the material loss of $\mathrm{Ag}$ can be reduced by the annealing process
so that the damping factor $\gamma$ is decreased.
Here the damping factor ratio $\Delta$ is included in order to tune the material
loss of $\mathrm{Ag}$ and is set to be $\Delta=0.1$ in the following analysis.
The permittivity of $\mathrm{Ti}_{3}\mathrm{O}_{5}$ is $\varepsilon_{d}=5.83$.
When the thickness of each layer is sufficiently small, i.e., $\left|ka\right|\ll1$,
where $k$ is the wave vector in the layer and $a$ is the thickness of the layer \cite{Elser2007APL},
the multilayer stack can be regarded as a homogeneous anisotropic effective medium
with a permittivity tensor based on the EMT
\begin{equation}
\label{eq:EMT}
\begin{aligned}
    &\varepsilon_{xy}^{\mathrm{EMT}}=f_{m}\varepsilon_{m}+f_{d}\varepsilon_{d}, \\
    &\varepsilon_{z}^{\mathrm{EMT}}=(f_{m}/\varepsilon_{m}+f_{d}/\varepsilon_{d})^{-1},
\end{aligned}
\end{equation}
which is related to the filling ratio of $\mathrm{Ag}$ and $\mathrm{Ti}_{3}\mathrm{O}_{5}$,
$f_{m}=a_{m}/(a_{m}+a_{d})$ and $f_{d}=a_{d}/(a_{m}+a_{d})$.
According to Eq.~(\ref{eq:EMT}), it is interesting that if the real part of the permittivity
of $\mathrm{Ag}$ is greater than that of the imaginary part, i.e.,
$\mathrm{Re}(\varepsilon_{m})\gg\mathrm{Im}(\varepsilon_{m})$,
the EMT-based ENZ wavelength (where $\mathrm{Re}(\varepsilon_{xy}^{\mathrm{EMT}})=0$)
and the EMT-base ENP wavelength (where the maximum of $\mathrm{Im}(\varepsilon_{z}^{\mathrm{EMT}})$
is obtained) can be approximately associated with the EMT-based filling ratio of $\mathrm{Ag}$
as $f_{m}^{\mathrm{EMT}}\approx \varepsilon_{d}/\left(\varepsilon_{d}-\mathrm{Re}(\varepsilon_{m})\right)$
and $f_{m}^{\mathrm{EMT}}\approx -\mathrm{Re}(\varepsilon_{m})
/\left(\varepsilon_{d}-\mathrm{Re}(\varepsilon_{m})\right)$, respectively.
Therefore, the EMT-based ENZ-ENP wavelength is related to the approximate condition of
$\varepsilon_{d}=-\mathrm{Re}(\varepsilon_{m})$, which implies that once the EMT-based
ENZ wavelength overlaps with the EMT-based ENP wavelength, the EMT-based filling ratio of
$\mathrm{Ag}$ should be $f_{m}^{\mathrm{EMT}}\approx1/2$.
Further algebraic calculation indicates that the accuracy result of the EMT-based filling
ratio of $\mathrm{Ag}$ is $f_{m}^{\mathrm{EMT}}=0.500$, with respect to the permittivity
of $\mathrm{Ag}$ in terms of Drude model and the permittivity of $\mathrm{Ti}_{3}\mathrm{O}_{5}$.
Correspondingly, the EMT-based effective permittivity tensor reads
$\varepsilon_{xy}^{\mathrm{EMT}}=0.000+0.012i$
and $\varepsilon_{z}^{\mathrm{EMT}}=7.238+2798.497i$,
presenting an extremely anisotropic permittivity property of the multilayer stack.
Regarding a TM-polarized light with non-vanishing $E_{x}$, $H_{y}$, and $E_{z}$ field
components propagating along the $z$-direction in the $x$-$z$ plane, the iso-frequency
contour (IFC) of the multilayer stack reads
\begin{equation}
\label{eq:EMT_IFC}
    k_{x}^{2}/\varepsilon_{z}^{\mathrm{EMT}} + k_{z}^{2}/\varepsilon_{xy}^{\mathrm{EMT}} = k_{0}^{2},
\end{equation}
while considering the extremely anisotropic permittivity property, i.e.,
$\left|\varepsilon_{z}^{\mathrm{EMT}}\right|\gg1\gg\left|\varepsilon_{xy}^{\mathrm{EMT}}\right|\approx0$
at the EMT-based ENZ-ENP wavelength, the IFC equation in Eq.~(\ref{eq:EMT_IFC}) can be approximately written as
\begin{equation}
\begin{aligned}
\label{eq:EMT_IFC_approx}
    k_{z}/k_{0} &= \sqrt{\varepsilon_{xy}^{\mathrm{EMT}}
        \left( 1-\left( k_{x}/k_{0} \right)^{2}/\varepsilon_{z}^{\mathrm{EMT}} \right)} \\
        &\approx \sqrt{\varepsilon_{xy}^{\mathrm{EMT}}}
        - \sqrt{\varepsilon_{xy}^{\mathrm{EMT}}}\left( k_{x}/k_{0} \right)^{2}
        /\left( 2\varepsilon_{z}^{\mathrm{EMT}} \right).
\end{aligned}
\end{equation}
Therefore, as shown in Fig.~1(b), according to Eq.~(\ref{eq:EMT_IFC_approx}), the IFC of the
multilayer stack at the EMT-based ENZ-ENP wavelength forms flat lines of the value
$k_{z}/k_{0}\approx(\varepsilon_{xy}^{\mathrm{EMT}})^{1/2}$ with respect to a wide $k_{x}/k_{0}$
wave vector range, which leads to a diffraction-free optical beam propagation in the
multilayer stack since all spatial components propagate with the same phase velocity along
the $z$-direction.
Furthermore, due to the fact that $\mathrm{Re}(\varepsilon_{xy}^{\mathrm{EMT}})=0$
and $\left|\varepsilon_{xy}^{\mathrm{EMT}}\right|\ll1$, the phase variation of the
diffraction-free optical beam propagation in the multilayer stack is close to zero.

However, previous studies show that the metal-dielectric multilayer stack possesses strong
optical nonlocality \cite{Elser2007APL,Orlov2011PRB,Chebykin2011PRB,Chebykin2012PRB},
leading to the effective permittivity tensor not only related to the frequency but also to
the wave vector, especially when the frequency approaches to the ENZ position of the structure.
Such optical nonlocality can be analyzed via the transfer-matrix method, in which the multilayer
stack is considered as a one-dimensional photonic crystal with the dispersion relation to the
TM-polarized light as
\begin{equation}
\begin{aligned}
\label{eq:nonloc}
    \cos( k_{z}(a_{m}+a_{d}) )
        &= \cos( k_{m}a_{m} )\cos( k_{d}a_{d} ) \\
        &- ( \varepsilon_{m}k_{d}/\varepsilon_{d}k_{m}
        + \varepsilon_{d}k_{m}/\varepsilon_{m}k_{d} )/2
        \sin( k_{m}a_{m} )\sin( k_{d}a_{d} ),
\end{aligned}
\end{equation}
with the wave vector $k_{i}=\sqrt{\varepsilon_{i}k_{0}^{2}-k_{x}^{2}}$ for $i=m,d$.
Accordingly, the effective permittivity component in $x$- and $y$-direction including the
optical nonlocality reads
\begin{equation}
\begin{aligned}
\label{eq:nonloc_eps}
    \varepsilon_{xy}^{\mathrm{nonloc}} &=
        \arccos^{2}[\cos(\sqrt{\varepsilon_{m}}k_{0}a_{m})\cos(\sqrt{\varepsilon_{d}}k_{0}a_{d}) \\
        &-(\sqrt{\varepsilon_{m}/\varepsilon_{d}}+\sqrt{\varepsilon_{d}/\varepsilon_{m}})/2
        \sin(\sqrt{\varepsilon_{m}}k_{0}a_{m})
        \sin(\sqrt{\varepsilon_{d}}k_{0}a_{d})]
        /(k_{0}^{2}(a_{m}+a_{d})^{2}).
\end{aligned}
\end{equation}
On the other hand, the nonlocal effective permittivity component in the $z$-direction can be analytically
obtained according to the effective permittivity definition
$\varepsilon_{z}^{\mathrm{nonloc}}=\langle D_{z}\rangle/\langle E_{z}\rangle$
based on the transfer-matrix method.
Due to the optical nonlocality, the nonlocal ENZ wavelength and the nonlocal ENP wavelength for
the multilayer stack will be different from the EMT-based ENZ-ENP wavelength.
Therefore, modification on the multilayer geometry is necessary for a certain period of $a_{m}+a_{d}$,
in order to exactly overlap the nonlocal ENZ wavelength with the nonlocal ENP wavelength.
Figure~1(c) shows an example about the nonlocal effective permittivity components for a multilayer
stack with the period of $a_{m}+a_{d}=20\,\mathrm{nm}$.
Compared with the EMT results, the nonlocal results clearly indicate the shift of the
ENZ-ENP wavelength from the EMT calculated position of $\lambda_{c}^{\mathrm{EMT}}=466.896\,\mathrm{nm}$
to the new position of $\lambda_{c}^{\mathrm{nonloc}}=468.424\,\mathrm{nm}$,
while the nonlocal effective permittivity components are
$\varepsilon_{xy}^{\mathrm{nonloc}}=0.000+0.012i$
and $\varepsilon_{z}^{\mathrm{nonloc}}=7.214+2800.879i$.
Furthermore, the filling ratio of $\mathrm{Ag}$ related to the optical nonlocality
(nonlocal filling ratio for short) also slightly changes to
$f_{m}^{\mathrm{nonloc}}=0.501$, with respect to the period of the multilayer stack.
In more detail, Figs.~2(a--b) individually display the variations of the filling ratio
of $\mathrm{Ag}$ and the ENZ-ENP wavelength with respect to different periods of the
multilayer stack caused by the optical nonlocality.
For comparison, the EMT results are also plotted.
It is clear that as the multilayer stack period gets larger, the nonlocal filling ratio of $\mathrm{Ag}$
increases and the nonlocal ENZ-ENP wavelength also shifts to a longer wavelength.

\subsection{Diffraction-free and near-zero phase variation propagation}
Based on the finite element method (FEM), the diffraction-free optical beam propagation with
near-zero phase variation is demonstrated according to the numerical simulation
in terms of two anti-phase TM-polarized Gaussian beams that are excited via the
stimulating port boundary condition with an initial electric field distribution
following the Gaussian distribution function as $\mathrm{exp}\left(-(x+c_{0})^{2}/w_{0}^{2}\right)$,
where $c_{0}$ determines the beam center and $w_{0}$ determines the waist size.
Otherwise, the scattering boundary condition is applied to prevent the unwanted scattering
from other boundaries of the simulation region.
The simulation results are presented as the amplitude of the electric field, normalized by the maximum value.
Figure~3 shows the calculated optical beam propagation of the Gaussian beams with ultra-narrow beam waists
of $w_{0}=25\,\mathrm{nm}$ in the homogeneous anisotropic effective medium at the EMT-based ENZ-ENP
wavelength of $\lambda_{c}^{\mathrm{EMT}}=466.896\,\mathrm{nm}$.
Besides, Fig.~4 shows the optical beam propagation results calculated in the $20\,\mathrm{nm}$-period
multilayer stack at the nonlocal ENZ-ENP wavelength $\lambda_{c}^{\mathrm{nonloc}}=468.424\,\mathrm{nm}$.
As shown in Figs.~3(a) and 4(a), it is clear that the deep-subwavelength Gaussian beams can
propagate over a long distance ($>2\lambda_{c}^{\mathrm{nonloc}}$) without any wave front distortion.
At the same time, near-zero phase variation is also obtained along the propagation direction.
The centre-to-centre distance of $120\,\mathrm{nm}$ for the two deep-subwavelength Gaussian
beams remains well-defined as the beams propagating across the multilayer stack from the
bottom to the top.
It is noted that the confinement and propagation of the deep-subwavelength Gaussian beam inside
the multilayer stack is entirely due to the unique extremely anisotropic permittivity property,
and the optical beam path is well determined by the beam launching position,
which is distinguished from the situation in a subwavelength optical waveguide \cite{He2013JOSAB},
where the optical mode is confined by the waveguide boundary.
Besides the straight optical beam propagation, the flow of light can be flexibly modeled through controlling
the local geometry of the multilayer stack.
For instance, the beam path can be manipulated by gradually varying the direction of multilayers,
since the direction of the optical beam propagation is always vertical to the multilayer interface.
The designed geometries for achieving $90^{\circ}$ and $180^{\circ}$ bending of deep-subwavelength
Gaussian beams and the simulation results are shown in Figs.~3(b--c) and 4(b--c),
for the effective medium and the multilayer stack, respectively.
It is noted that for the effective medium the anisotropic effective permittivity tensor
is transformed according to the geometric relation between the global $xyz$-coordinates
and the local $vuw$-coordinates based on the transfermation optics as
\begin{equation}
\label{eq:trans_opt}
\stackrel{\leftrightarrow}{\varepsilon}(x,z) =
\begin{pmatrix}
    \varepsilon_{u}\cos^{2}\theta+\varepsilon_{w}\sin^{2}\theta &
    (\varepsilon_{u}-\varepsilon_{w})\sin\theta\cos\theta \\
    (\varepsilon_{u}-\varepsilon_{w})\sin\theta\cos\theta &
    \varepsilon_{u}\sin^{2}\theta+\varepsilon_{w}\cos^{2}\theta
\end{pmatrix}.
\end{equation}
The results indicate that the flow of light can indeed be manipulated while maintaining
the diffraction-free propagation with near-zero phase variation.
For comparison, Fig.~4(d) presents the optical beam propagation at the wavelength
of $\lambda=464\,\mathrm{nm}$, which is off the nonlocal ENZ-ENP wavelength of the
$20\,\mathrm{nm}$-period multilayer stack, while the corresponding nonlocal effective
permittivity components read
$\left.\varepsilon_{xy}^{\mathrm{nonloc}}\right|_{\lambda=464\,\mathrm{nm}} = 0.121+0.012i$ and
$\left.\varepsilon_{z}^{\mathrm{nonloc}}\right|_{\lambda=464\,\mathrm{nm}} = -336.266+42.895i$.
It is clear that by overlapping the ENZ wavelength and the ENP wavelength,
the multilayer stack with such an extremely anisotropic effective permittivity
tensor supports diffraction-free propagation with near-zero phase variation.
Also, it implies that the diffraction-free propagation supported by the anisotropic
effective permittivity tensor is not extremely sensitive to the value of wavelength,
but for the near-zero phase variation, the ENZ wavelength must be satisfied.

In the previous analysis, the damping factor ratio of $\mathrm{Ag}$ is set as $\Delta=0.1$
in order to maintain a small material loss to extend the propagation distance of the Gaussian beams.
In fact, the variation of the damping factor in the Drude model will affect the optical beam propagation
loss and the phase variation.
Figure~5(a) represents the variation of the propagation length and the effective refractive index
along the propagation direction with respect to different damping factor ratios of $\mathrm{Ag}$.
The effective refractive index is defined based on the nonlocal effective permittivity component
as $n_{z}=(\varepsilon_{xy}^{\mathrm{nonloc}})^{1/2}$ according to Eq.~(\ref{eq:nonloc_eps}).
The real part of the effective refractive index $\mathrm{Re}(n_{z})$ is proportional to the phase
variation of the propagating beam, while the imaginary part $\mathrm{Im}(n_{z})$ is inversely
proportional to the propagation length of the beam as $\lambda/(4\pi\mathrm{Im}(n_{z}))$,
where $\lambda$ is the wavelength of the beam in free space.
It is clear that as the damping factor ratio of $\mathrm{Ag}$ gets larger, the real part of
the effective refractive index increases, leading to an obvious beam phase variation along
the propagation direction.
On the other hand, the propagation length also decreases as the material loss gets higher.
Figures~5(b--d) show the propagation of the two anti-phase Gaussian beams in the $20\,\mathrm{nm}$-period
multilayer stack with different damping factor ratio of $\Delta=0.1$, $0.5$, and $1.0$ at the nonlocal
ENZ-ENP wavelengths, illustrating the dependences of propagation length and phase variation on the damping
factor ratio of $\mathrm{Ag}$.
It is shown that a smaller material loss will benefit the diffraction-free and near-zero phase
variation optical beam propagation in the multilayer stack.

Finally, it is worth mentioning that according to the previous experimental
work of optical hyperlens \cite{Liu2007Sci},
our above theoretical analysis of extremely anisotropic metamaterials can be
realized in experiments.
The ultra-narrow Gaussian beams can be achieved by inscribing nanometer scale slots
or holes into a thin chrome layer coated on the top of the sample surface,
while the impedance matching substrate can be applied at the bottom of the multilayer
stack in order to reduce the reflection caused by the mismatched impedance.

\section{Conclusions}

To conclude, extremely anisotropic metal-dielectric multilayer metamaterials with identical ENZ wavelength
and ENP wavelength have been designed based on the transfer-matrix method including the optical nonlocality.
Such metamaterial possesses a near-flat IFC with wave vector values close to zero over a broad wave vector range.
This unique property is utilized to obtain the diffraction-free deep-subwavelength optical beam propagation
with near-zero phase variation, and it is demonstrated that the optical beam propagation can be manipulated flexibly
by tuning the direction of the multilayer stack.
The influences of multilayer period and material loss on the optical beam propagation are also studied.
The current study is of great potentials in many applications such as optical imaging, optical integration,
and on-chip optical communication.

\section*{Acknowledgements}

This work was supported by the Intelligent Systems Center and the Energy Research and Development Center
at Missouri S\&T, the University of Missouri Interdisciplinary Intercampus Research Program,
the Ralph E. Powe Junior Faculty Enhancement Award, and the National Science Foundation under
grant CBET-1402743.


\newpage                  %
\section*{Figure Captions}%

\textbf{FIG.~1}. (Colour online)
(a) Schematic of $\mathrm{Ag}$-$\mathrm{Ti}_{3}\mathrm{O}_{5}$ multilayer stack.
(b) The IFC of the $\mathrm{Ag}$-$\mathrm{Ti}_{3}\mathrm{O}_{5}$
multilayer stack at the EMT-based ENZ-ENP wavelength,
where the wave vector $k_{z}/k_{0}$  possesses near-zero flat real part (red lines) and
imaginary part (black lines), compared with the IFC of air (green circle).
(c) The EMT-based (dashed curves) and the nonlocal (solid curves) effective permittivity
components, with the marked EMT-based (dashed vertical line) and the nonlocal
(solid vertical line) ENZ-ENP wavelength.

\vspace{5.0mm}
\noindent
\textbf{FIG.~2}. (Colour online)
The variation of (a) the filling ratio of $\mathrm{Ag}$ and
(b) the ENZ-ENP wavelength with respect to different periods of the
$\mathrm{Ag}$-$\mathrm{Ti}_{3}\mathrm{O}_{5}$
multilayer stack with optical nonlocality, compared with the EMT-based
results representing the zero-period case.

\vspace{5.0mm}
\noindent
\textbf{FIG.~3}. (Colour online)
The diffraction-free optical beam propagation with near-zero phase variation of
two anti-phase deep-subwavelength TM-polarized Gaussian beams in the effective
medium for the case of
(a) straight propagation,
(b) $90^{\circ}$ beam bending, and
(c) $180^{\circ}$ beam bending at the EMT-based ENZ-ENP wavelength.

\vspace{5.0mm}
\noindent
\textbf{FIG.~4}. (Colour online)
The diffraction-free optical beam propagation with near-zero phase variation of two Gaussian
beams in the $\mathrm{Ag}$-$\mathrm{Ti}_{3}\mathrm{O}_{5}$
multilayer stack for the case of
(a) straight propagation,
(b) $90^{\circ}$ beam bending, and
(c) $180^{\circ}$ beam bending at the nonlocal-based ENZ-ENP wavelength.
(d) The optical beam propagation of two Gaussian beams off the nonlocal-based
ENZ-ENP wavelength.

\vspace{5.0mm}
\noindent
\textbf{FIG.~5}. (Colour online)
(a) The variation of the propagation length (blue curve) and the real part
of the effective refractive index (red curve) with respect to different
damping factor ratios of $\mathrm{Ag}$.
(b--d) The diffraction-free optical beam propagation with near-zero phase variation
of two Gaussian beams in the $\mathrm{Ag}$-$\mathrm{Ti}_{3}\mathrm{O}_{5}$
multilayer stack with respect to the damping factor ratio of
(b) $\Delta=0.1$, (c) $\Delta=0.5$, and (d) $\Delta=1.0$ at the nonlocal-based
ENZ-ENP wavelengths.

\newpage
\begin{figure}[htb]
    \centering
    \includegraphics[width=8.5cm]{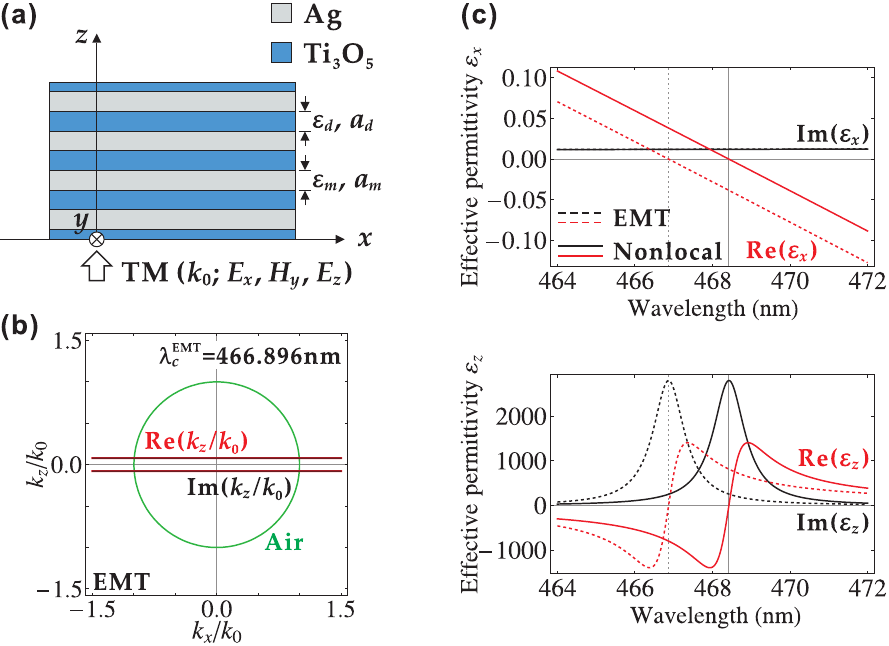}
    \caption{Lei~Sun, Xiaodong~Yang, Wei~Wang, and Jie~Gao}
\end{figure}

\newpage
\begin{figure}[htb]
    \centering
    \includegraphics[width=8.0cm]{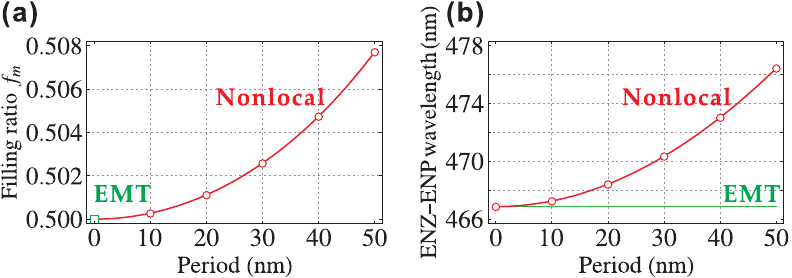}
    \caption{Lei~Sun, Xiaodong~Yang, Wei~Wang, and Jie~Gao}
\end{figure}

\newpage
\begin{figure}[htb]
    \centering
    \includegraphics[width=8.0cm]{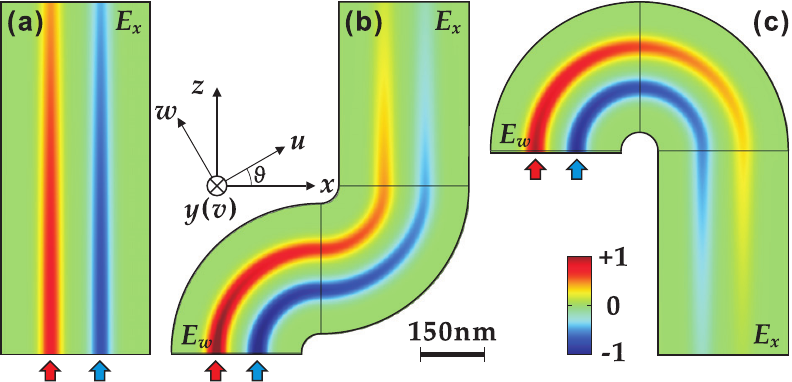}
    \caption{Lei~Sun, Xiaodong~Yang, Wei~Wang, and Jie~Gao}
\end{figure}

\newpage
\begin{figure}[htb]
    \centering
    \includegraphics[width=9.8cm]{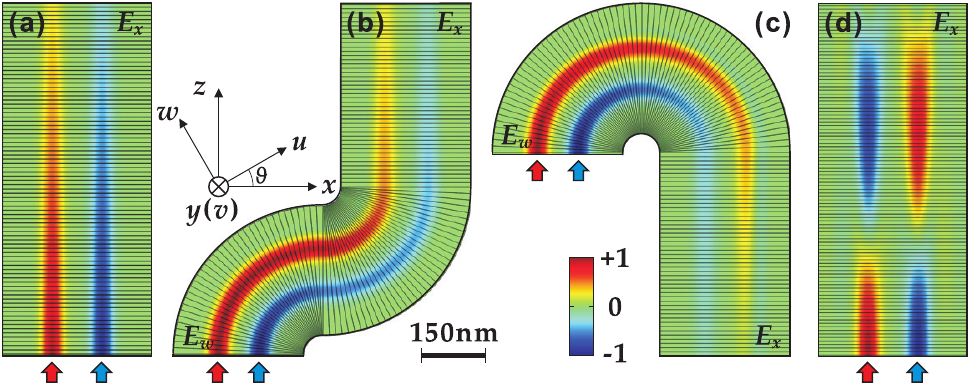}
    \caption{Lei~Sun, Xiaodong~Yang, Wei~Wang, and Jie~Gao}
\end{figure}

\newpage
\begin{figure}[htb]
    \centering
    \includegraphics[width=6.5cm]{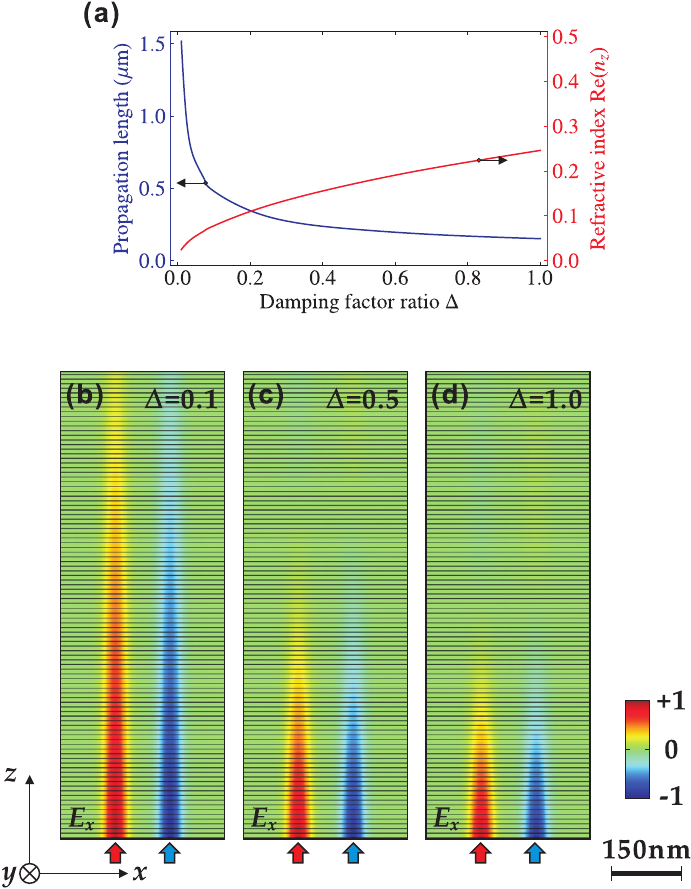}
    \caption{Lei~Sun, Xiaodong~Yang, Wei~Wang, and Jie~Gao}
\end{figure}

\end{document}